\documentclass[aps,prl]{revtex4}
\usepackage{epsfig,pstricks}
\usepackage{amsmath,amssymb}

\def\CC{{\rm\kern.24em \vrule width.04em height1.46ex depth-.07ex
\kern-.30em C}}
\def\RR{{\rm
         \vrule width.04em height1.58ex depth-.0ex
         \kern-.04em R}}
\def\id{{\rm 1\kern-.22em l}}

\newcommand{\beq}{\begin{equation}}
\newcommand{\beqa}{\begin{eqnarray}}
\newcommand{\nbeqa}{\begin{eqnarray*}}
\newcommand{\eeq}{\end{equation}}
\newcommand{\eeqa}{\end{eqnarray}}
\newcommand{\neeqa}{\end{eqnarray*}}
\newcommand{\bra}[1]{\left\langle #1 \right |}
\newcommand{\ket}[1]{\left | #1 \right\rangle}

\newcommand{\expect}[1]{\left\langle #1 \right\rangle}
\newcommand{\dexpect}[1]{\left\langle\hspace{-2mm}\left\langle\, 
		#1 \,\right\rangle\hspace{-2mm}\right\rangle}
\newcommand{\diag}{{\rm diag}\;}
\newcommand{\bigfrac}[2]{\mbox {${\displaystyle \frac{ #1 }{ #2 }}$}}

\begin{document}
\tighten

\title{Constructing $N$-qubit entanglement monotones 
       from antilinear operators}
\author{Andreas Osterloh$^{1,2}$}
\author{Jens Siewert$^{1,3}$}
\affiliation{$^1$ MATIS-INFM $\&$ Dipartimento di Metodologie Fisiche e
    Chimiche (DMFCI), viale A. Doria 6, 95125 Catania, Italy\\
             $^2$ Institut f\"ur Theoretische Physik, 
                  Universit\"at Hannover, D-30167 Hannover, Germany\\
             $^3$ Institut f\"ur Theoretische Physik, 
                  Universit\"at Regensburg, D-93040 Regensburg, Germany}

\begin{abstract}
We present a method to construct entanglement measures 
for pure states of multipartite qubit systems. 
The key element of our approach
is an antilinear operator that we call {\em comb} in reference to the
{\em hairy-ball theorem}. 
For qubits (or spin 1/2) the combs are automatically 
invariant under $SL(2,\CC)$.
This implies that the {\em filters} obtained from the 
combs are entanglement monotones by construction.
We give alternative formulae for the concurrence and 
the 3-tangle as expectation values of certain antilinear operators.
As an application we discuss inequivalent
types of genuine four-qubit entanglement.

\end{abstract}

\maketitle

Entanglement is one the most striking features of quantum mechanics,
but it is also one of its most counterintuitive
consequences of which we still have rather incomplete knowledge~\cite{Bell87}.
Although the concentrated effort during the past decade has produced
an impressive progress, there is no general
qualitative and quantitative theory of entanglement.

A pure quantum-mechanical state of distinguishable particles is 
called disentangled with respect to a given 
partition ${\cal P}$ of the system {\em iff} it can be written as a 
tensor product of the parts of this partition. In the opposite case,
the state must contain some finite amount of entanglement. 
The question then is
to characterize and quantify this entanglement.

As to {\em measuring} the amount of entanglement in a given pure
multipartite state, the first major step was made by
Bennett {\it et al.}~\cite{BennettDiVincenzo96} who discovered that the partial
entropy of a party in a bipartite quantum state is a measure of
entanglement. It coincides (asymptotically) with the entanglement of formation
(i.e., the number of Einstein-Podolsky-Rosen pairs required to prepare
a given state).
Subsequently, the entanglement of formation of a two-qubit state
was related to the concurrence~\cite{Hill97,Wootters98}.
Interestingly, by 
exploiting the knowledge of the mixed-state concurrence, a measure
for three-partite pure states could be derived, the so-called
3-tangle $\tau_3$~\cite{Coffman00}. In terms of the coefficients of the 
wavefunction $\{\psi_{000},\psi_{001},\ldots,\psi_{111}\}$
in the standard basis it reads
\nbeqa
\tau_3 &=& |d_1 - 2d_2 + 4d_3|\\
  d_1&=& \psi^2_{000}\psi^2_{111} + \psi^2_{001}\psi^2_{110} + \psi^2_{010}\psi^2_{101}+ \psi^2_{100}\psi^2_{011} \\
  d_2&=& \psi_{000}\psi_{111}\psi_{011}\psi_{100} + \psi_{000}\psi_{111}\psi_{101}\psi_{010}\\ 
    &&+ \psi_{000}\psi_{111}\psi_{110}\psi_{001} + \psi_{011}\psi_{100}\psi_{101}\psi_{010}\\
    &&+ \psi_{011}\psi_{100}\psi_{110}\psi_{001} + \psi_{101}\psi_{010}\psi_{110}\psi_{001}\\
  d_3&=& \psi_{000}\psi_{110}\psi_{101}\psi_{011} + \psi_{111}\psi_{001}\psi_{010}\psi_{100}\ \ .
\neeqa
This was a remarkable step since, loosely speaking, it opened the path 
to studying multipartite entanglement on solid grounds. 
Further, it was noticed by Uhlmann that antilinearity is
an important property of operators that measure 
entanglement~\cite{Uhlmann}.
A particularly interesting consequence of the 3-tangle formula
was presented by D\"ur {\it et al.} who found that there are two inequivalent
classes of states with three-party entanglement~\cite{Duer00}.

Another important aspect of the research on entanglement measures 
was the question regarding the requirements for a function
that represents an entanglement monotone~\cite{MONOTONES}. It turned out
that the essential property to be satisfied is non-increasing behavior
on average under stochastic local operations and classical 
communication (SLOCC)~\cite{SLOCC,Duer00}.
Later, Verstraete {\em et al.} have demonstrated that, in general,
an entanglement monotone can be obtained from any homogeneous positive
function of pure-state density matrices that remain invariant
under determinant-one SLOCC operations~\cite{VerstraeteDM03}.

Despite the enormous effort, the only truly operational entanglement measure
for arbitrary mixed states at hand, up to now, is the concurrence.
For pure states we have a slightly farther view up to 
systems of two qutrits~\cite{Cereceda03,Briand03},
and for three qubits, due to the  3-tangle.
Various multipartite measures for pure-state entanglement have been proposed.
However, none of them yields zero for all possible
product states, see, e.g., Refs.~\cite{Barnum03,HEYDARI04,Wallach,Wong00}.  

This motivated the quest for an operational
entanglement measure based on the requirement that it be zero for
product states (not only for completely separable pure states).
In particular, the goal has been to explore the idea that entanglement
monotones are related to antilinear operators as pointed out for the
concurrence by Uhlmann~\cite{Uhlmann}.
Here we show that it is possible to construct a {\em filter},
i.e., an operator that has zero expectation value for all product states. 
It will turn out that these filters are entanglement monotones 
by construction. Interestingly, the two-qubit concurrence
and the 3-tangle have various equivalent filter representations (see below).
In order to illustrate the application of  the method to a
nontrivial example, we will present filters for
four-qubit states that are able to distinguish inequivalent 
types of genuine four-qubit entanglement. We use the term 
{\em genuine $N$-qubit entanglement} in a more restricted sense
than, e.g., in Refs.~\cite{Duer00,Toth2005}: a state with only
genuine $N$-partite entanglement does not contain any genuine $(N-k)$-partite
entanglement (or ``sub-tangle'') with $1\le k\le N-2$. 
In this sense the only class of 
three-qubit states with genuine three-partite entanglement is represented
by the GHZ state.

{\em Combs and filters -- }
The basic concept is that of the {\em comb}, i.e.,
an antilinear operator $A$ with 
zero expectation value for all states of a certain Hilbert space
${\cal H}$. 
That is,
\begin{equation}\label{def:Toeter}
\bra{\psi} A \ket{\psi}=\bra{\psi} L C\ket{\psi}
	=\bra{\psi} L \ket{\psi^*}\equiv 0
\end{equation}
for all $\ket{\psi}\in{\cal H}$, where $L$ is a linear operator and 
$C$ is the complex conjugation.
Here $A$ necessarily has to be antilinear 
(a linear operator with this property is zero itself).
For simplicity we abbreviate
\begin{equation}
\bra{\psi} L C\ket{\psi}=:\expect{L}_C \ \ .
\label{defexpect}
\end{equation}
Note that the complex conjugation is {\em included}
in the definition of the expectation value $\langle \ldots \rangle_C$
in Eq.\ (\ref{defexpect}).

We will use the comb operators~\footnote{
                If the antilinear operator $A=L C$ is 
                a comb (with the complex conjugation $C$),
                for the sake of brevity
                we will also call the linear operator $L$
                a comb.}
in order to construct the desired filters.
It is worth mentioning already at this point 
that such a filter is invariant under 
${\cal P}$-local unitary 
transformations if the 
combs have this property. 
Even more, it is invariant under the complex 
extension of the corresponding unitary group 
which is isomorphic to
the special linear group. 
Since the latter represents the 
SLOCC operations for qubits~\cite{SLOCC,Duer00},
the filters will be entanglement monotones by construction.

In this work, we restrict our focus to multipartite systems of
qubits (i.e., spin 1/2). The local Hilbert space is 
${\cal H}_j=\CC^2=:{\frak h}$ for all $j$. We need the Pauli matrices
$\sigma_0:=\id$, $\sigma_1:=\sigma_x$,
$\sigma_2:=\sigma_y$, and
$\sigma_3:=\sigma_z$.
A simple example for a single-qubit comb
is the operator 
$\sigma_y$:
\[
\bra{\psi} \sigma_y C\ket{\psi}=\expect{\sigma_y}_C\equiv 0 \ \ .
\]
Notice that any tensor product
$f(\{\sigma_\mu\}):=\sigma_{\mu_1}\otimes\dots\otimes\sigma_{\mu_n}$ 
with an odd number $N_y$ of $\sigma_y$ is an $n$-site comb. 
This can be seen immediately from
$\expect{f(\{\sigma_\mu\})}_C \equiv
\bra{\psi} f(\{\sigma_\mu\}) \ket{\psi^*}$ 
$=\left (\bra{\psi^*} f(\{\sigma_\mu^*\}) \ket{\psi}\right )^*$
$= (-1)^{N_y} \bra{\psi} f(\{\sigma_\mu\})^\dagger \ket{\psi^*}$ 
$=(-1)^{N_y}\expect{f(\{\sigma_\mu\})}_C$.
Since its expectation value is a bi(anti-)linear expression in the
coefficients of the state we will denote it a comb
of {\em order 1}. In general we will call a comb to
be {\em of order n} if its expectation value is $2n$-linear
in the coefficients of the state.
There is another single-qubit comb which is of 2nd order. 
One can verify that for an arbitrary single-qubit state
\begin{equation}
0=\expect{\sigma_\mu}_C\expect{\sigma^\mu}_C:=
\sum_{\mu,\nu=0}^3 
\expect{\sigma_\mu}_C
		g^{\mu,\nu}\expect{\sigma_\nu}_C\; ,
\end{equation}
with
$g^{\mu,\nu}=\diag\{-1,1,0,1\}$
being very similar to the Minkowski metric.

It will prove useful to introduce the embedding
\begin{equation}\label{embedding}
{\cal E}_n\; :\; 
\begin{array}{ccc}
{\cal H}   &\hookrightarrow& \mathfrak{H}_n={\cal H}^{\otimes\; n} \\
\ket{\psi} &\longrightarrow& {\cal E}_n\ket{\psi} = \ket{\psi}^{\otimes\; n}
\ \ .
\end{array}
\end{equation}
Further define the product $\bullet$ for operators 
$O$, $P$: ${\cal H} \longrightarrow {\cal H}$ such that
\begin{equation}\label{embedding:product}
O \bullet P\; :\; 
\begin{array}{ccc}
\mathfrak{H}_2   &\rightarrow& \mathfrak{H}_2 \\
O \bullet P {\cal E}_2(\ket{\psi}) &=& O\ket{\psi} \otimes P\ket{\psi}
\ \ .
\end{array}
\end{equation}
Then we have the single-site (${\cal H}=\CC^2$) 
comb $\sigma_y$ in $\mathfrak{H}_1={\cal H}$
and $\sigma_\mu\bullet\sigma^\mu$ in $\mathfrak{H}_2$.
We will discuss in more detail below how one can see that
both $\sigma_y $ and $\sigma_\mu\bullet\sigma^\mu$ are 
invariant under SLOCC.

With these two one-site combs we are now equipped to construct
filters for multipartite qubit systems. For $n$-qubit filters we will 
use the symbol ${\cal F}^{(n)}$. 
For two qubits the filters are
\begin{eqnarray}\label{2-filters}
{\cal F}^{(2)}_1\ &=&\ \sigma_y\otimes \sigma_y \\
{\cal F}^{(2)}_2\ &=&\ \frac{1}{3}\ (\sigma_\mu\otimes\sigma_\nu)\bullet
		  (\sigma^\mu\otimes \sigma^\nu)\ \ .
\end{eqnarray}
Both forms are explicitly permutation invariant, and they are filters
since, if the state were a product, the combs would annihilate
its expectation value.
From the filters we obtain the pure-state concurrence in two different
equivalent forms:
\begin{eqnarray}\label{2-measures}
C\ &=&\ \left |\dexpect{{\cal F}^{(2)}_1}_C\right| \\
{\hspace*{-2mm}C}^2\ &=&\  \left|\dexpect{{\cal F}^{(2)}_2}_C\right|
\ \equiv \ \frac{1}{3}\ \left |\expect{\sigma_\mu\otimes\sigma_\nu}_C
    \expect{\sigma^\mu\otimes\sigma^\nu}_C\right| \ \ . \nonumber
\end{eqnarray}
While the first form in Eq.\ (\ref{2-measures}) 
has the well-know
convex-roof extension of the pure-state concurrence via the 
matrix~\cite{Hill97,Wootters98,Uhlmann} 
\begin{equation}\label{mixed1}
R \ = \ \sqrt{\rho}\ \sigma_y\otimes\sigma_y\  
                           \rho^*\  \sigma_y\otimes\sigma_y \
    \sqrt{\rho}
\end{equation}
it can be shown that the second form in Eq.\ (\ref{2-measures}) 
leads to 
\begin{eqnarray}\label{mixed2}
   R^2 &=& \sqrt{\rho}\ \sigma_\mu\otimes\sigma_\nu \ \rho^*\  
                        \sigma_\kappa\otimes\sigma_\lambda\\
        &&\quad  \rho\ \sigma^\mu\otimes\sigma^\nu\  \rho^*\  
                       \sigma^\kappa\otimes\sigma^\lambda
          \ \sqrt{\rho} \ \ .\nonumber
\end{eqnarray}

Now let us consider the 3-tangle~\cite{Coffman00}.
For states of three qubits we find, e.g.,
\begin{eqnarray}\label{3-filters}
{\cal F}^{(3)}_1 &\ =&\ (\sigma_\mu\otimes\sigma_y\otimes\sigma_y)\bullet
		  (\sigma^\mu\otimes \sigma_y\otimes\sigma_y)\\
{\cal F}^{(3)}_2 &\ =&\ \bigfrac{1}{3}\
		(\sigma_\mu\otimes\sigma_\nu\otimes\sigma_\lambda)\bullet
		  (\sigma^\mu\otimes \sigma^\nu\otimes\sigma^\lambda)
 \ \ . \ \ 
\end{eqnarray}
Both ${\cal F}^{(3)}_1$ and ${\cal F}^{(3)}_2$
are filters and the latter is explicitly permutation invariant. 
From these operators the pure-state 3-tangle is obtained in the following way:
\begin{eqnarray}\label{3-measures}
\tau_3&=&\left |\dexpect{{\cal F}^{(3)}_1}_C\right| 
      = \left|\dexpect{{\cal F}^{(3)}_2}_C\right|
\end{eqnarray}
Interestingly, {\em all} three-qubit
filters reproduce the 3-tangle as entanglement measure.
We mention, however, that there is no immediate extension to mixed states 
as in the case of the 'alternative' two-qubit concurrence, 
\mbox{Eq.\ (\ref{mixed2})}.

{\em Invariance of filters under SLOCC -- }
Up to here we have shown that the concepts of combs and filters 
reproduce the well-known pure-state entanglement measures 
of concurrence and
3-tangle. Now we will briefly explain that the expectation values
of $N$-qubit filter operators are invariant under SLOCC
operations.

It has been demonstrated in Refs.~\cite{SLOCC,Duer00,VerstraeteDM03} 
that invariance under SLOCC operations reduces to invariance with
respect to the group $SL(2,\CC)^{\otimes N}$. As to the 
single-qubit combs, it is easily verified that 
$V\sigma_yV^T=\sigma_y$ for any local transformation $V\in SL(2,\CC)$.
Further, there is an operator identity
     $(V\otimes V)(\sigma_{\mu} \otimes \sigma^{\mu})(V^T\otimes V^T)
       = \sigma_{\mu} \otimes \sigma^{\mu}$ which 
expresses directly the $SL(2,\CC)$ invariance of 
$\sigma_\mu\bullet\sigma^\mu$.

We can, therefore, conclude that
tensor products of single-site combs are invariant with respect to
$SL(2,\CC)$ operations at each site. 
Hence, as long as an $N$-qubit filter operator is built from tensor products
of single-site combs it will be invariant under SLOCC operations.

Note that the SLOCC invariance of a filter expectation value 
with respect to some state 
(which vanishes if there is a way of writing the state as a tensor product)
means that this expectation value represents an entanglement 
monotone~\cite{MONOTONES,SLOCC}.
This imposes the question {\em what kind} of
entanglement is being measured by these quantities. Recall the case
of three qubits~\cite{Duer00} where there are two kinds of 
three-partite entanglement: GHZ-type (or genuine) entanglement which is
detected by the 3-tangle and {\em W}-type entanglement with zero 3-tangle.
As to the $N$-qubit case, we do not have a complete answer to the 
question above at this moment (although it is straightforward to
write down $N$-qubit filter operators). There are, however, indications
that the filters measure the {\em maximally entangled states}
discussed, e.g., in Refs.~\cite{Gisin98,VerstraeteDM03}.
In order to see this it is instructive to consider the 
four-qubit case.

{\em Filters for four-qubit states -- }
Classifications of four-qubit states with respect
to their entanglement properties have been studied, e.g., in 
Refs.~\cite{VerstraeteDMV02,BriandLT03,Myake03}. Here we  introduce
several four-qubit filter operators and study the classes of entangled
states they are measuring.

A four-qubit filter has the property that its expectation value
for a given state is zero if the state is separable, i.e.,
if there is a one-qubit or a two-qubit part which can be factored out
(note that for a three-qubit filter it is enough to extract
one-qubit parts only). An expression that obeys this requirement
for any single qubit and any combination of qubit pairs is given by
\begin{equation}
\label{fourbit6lin}
{\cal F}^{(4)}_1  = 
                (\sigma_\mu\sigma_\nu\sigma_y\sigma_y)\bullet
                  (\sigma^\mu\sigma_y\sigma_\lambda\sigma_y)
        \bullet(\sigma_y\sigma^\nu\sigma^\lambda\sigma_y) \ \ .
\end{equation}
Recall that any combination of the type 
$\sigma_{\mu}\sigma_y$ ($\mu\ne 2$) 
represents a two-qubit comb.
Note that the expectation value of an $n$th-order four-qubit 
     filter has to be taken with
     respect to the corresponding ${\mathfrak H}_n$, 
     see Eq.\ (\ref{embedding}). 
It is straightforward to check that for a four-qubit GHZ state
\begin{equation}
\label{ghz4}
    \ket{\Phi_1}\ =\ \frac{1}{\sqrt{2}}( \ket{0000}  +  \ket{1111})
\end{equation}
we have  $\bra{\Phi_1}{\cal F}^{(4)}_1 \ket{\Phi_1^*}=1$. 
However, there is another state for which
$\langle{\cal F}^{(4)}_1 \rangle_C$ does not vanish. For
\begin{equation}
\label{wlength6}
    \ket{\Phi_2}\ =\ \frac{1}{\sqrt{6}}
    (\sqrt{2}\ket{1111}+\ket{1000}+\ket{0100}+\ket{0010}+\ket{0001})
\end{equation}
we find  $\bra{\Phi_2}{\cal F}^{(4)}_1 \ket{\Phi_2^*}=8/9$.

Besides the 3rd-order filter ${\cal F}^{(4)}_1$ there exist also
filters of 4th order  and of 6th order. Examples are
\begin{eqnarray}\label{4-filters}
{\cal F}^{(4)}_2 &=&
                (\sigma_\mu\sigma_\nu\sigma_y\sigma_y)\bullet
                  (\sigma^\mu\sigma_y\sigma_\lambda\sigma_y)\bullet
   \nonumber\\
   &&\qquad\qquad \bullet        (\sigma_y\sigma^\nu\sigma_y\sigma_\tau) 
            \bullet
                (\sigma_y\sigma_y\sigma^\lambda\sigma^\tau)\\
{\cal F}^{(4)}_3 &=&\bigfrac{1}{2}
                (\sigma_\mu\sigma_\nu\sigma_y\sigma_y)\bullet
                  (\sigma^\mu\sigma^\nu\sigma_y\sigma_y)
\bullet(\sigma_\rho\sigma_y\sigma_\tau\sigma_y) \bullet
         \nonumber\\ &&\qquad  \bullet    
          (\sigma^\rho\sigma_y\sigma^\tau\sigma_y)
 \bullet(\sigma_y\sigma_\rho\sigma_\tau\sigma_y) \bullet
                (\sigma_y\sigma^\rho\sigma^\tau\sigma_y)
 \ \ . \nonumber
\end{eqnarray}
While ${\cal F}^{(4)}_2$ measures only GHZ-type entanglement
($\bra{\Phi_1}{\cal F}^{(4)}_2 \ket{\Phi_1^*}=1$)
the 6th-order filter ${\cal F}^{(4)}_3$ has the non-zero expectation values 
1/2 for the GHZ state and 1 for yet another state,
\begin{equation}
\label{wlength4}
    \ket{\Phi_3}\ =\ \frac{1}{2}
    (\ket{1111}+\ket{1100}+\ket{0010}+\ket{0001})\ \ .
\end{equation}
${\cal F}^{(4)}_1$ has zero expectation value for this state.
Finally, it is not difficult to convince oneself that all
four-qubit filters ${\cal F}_j^{(4)}$ ($j=1,2,3$) have zero expectation 
value for the
four-qubit {\em W} state $1/2(\ket{0111}+\ket{1011}+\ket{1101}+\ket{1110})$.

By analyzing the states $\ket{\Phi_j}$ 
we find that they are peculiar in the sense that the local density
operators for each qubit are given by $\frac{1}{2}\id$. As also all other
reduced
density operators do not have any $k$-tangle \mbox{($2\le k\le3$)} we conclude
that the $\ket{\Phi_j}$ represent classes of genuine four-qubit
entanglement. They are maximally entangled in the sense of 
Refs.~\cite{Gisin98,VerstraeteDM03}. 
Note that they
cannot be transformed into one another by SLOCC operations:
A state with a finite expectation value for one filter cannot be
transformed by means of SLOCC operations into a state with zero expectation
value for the same filter.
For example, ${\cal F}^{(4)}_2$ detects the GHZ state $\ket{\Phi_1}$
but gives zero for the other two states. Therefore, the four-qubit
entanglement in those states must be different from that of the GHZ state.

Hence, there are at least three inequivalent types of genuine entanglement
for four qubits~\footnote{In fact, there are {\em exactly} three 
                         maximally entangled states for four qubits. 
                         This will be discussed in a forthcoming publication.}.
We mention that the three maximally entangled states $\ket{\Phi_j}$
are not distinguished by the classification for pure four-qubit states
of Ref.~\cite{VerstraeteDMV02}. This can be seen by computing 
the expectation values of the four-qubit filters and the
reduced one-qubit density matrices for each of the nine class representatives
of Ref.~\cite{VerstraeteDMV02}. Only the classes 1--4 and 6 have 
non-vanishing ``4-tangle''. The corresponding local density matrices
can be completely mixed {\em only} for class 1. Therefore, all three
states $\ket{\Phi_j}$ must belong to that class. 

{\em Conclusions -- }
We have presented a new and efficient way
of generating entanglement monotones. It is based on
operators which we called {\em filters}.
The expectation values of these operators are zero for
all possible  product states, not only for the completely factoring case.
The building blocks of the filters (denoted
{\em combs}) guarantee invariance under $SL(2,\CC)^{\otimes N}$ for qubits. As
a consequence, all filters are automatically entanglement monotones.
They are measures of genuine multipartite entanglement.
This circumvents the difficult task to construct entanglement 
monotones from the essentially known (linear) local unitary invariants.

Further advantages of our approach lie in the feasibility of constructing
specific monotones that vanish for certain separable (pure) states
and in the applicability of this concept to 
partitions into subsystems other than qubits (i.e., qutrits\ldots).
Although only filters for four qubits are given explicitly in this work,
it is possible to build filters for any $N>4$ from the two presented 
single-qubit combs in a straightforward manner.

As an immediate result of our method the concurrence 
for pure two-qubit states is reproduced.
Moreover, we have found an alternative expression for 
the concurrence with the corresponding convex roof extension.
The application of the method to pure three-qubit states
yields several operator-based expressions for the 3-tangle,
including an explicitly permutation-invariant form.
Finally we have given explicit expressions for four-qubit entanglement
measures that detect three different types of genuine four-qubit
entanglement. As we have found, these types of genuine four-qubit
entanglement are not distinguished by the classification 
of four-qubit states in Ref.~\cite{VerstraeteDMV02}. 

As to $N$-qubit systems, there remain various interesting questions.
Clearly, it would be desirable to have a recipe how to 
build invariant combs for more complicated systems (e.g. higher spin).
It would also be interesting to know what characterizes
a complete set of filters for a given $N$.
While it is not obvious how the convex roof construction for two qubits
can be generalized, we believe that the operator form of the
$N$-tangles in terms of filters makes it easier to solve this problem.
The question is whether there is a systematic way to obtain a
convex-roof construction for a given filter with general
multi-linearity.

Returning to the case of two qubits, one may conclude that 
the crucial quality of $\sigma_y\otimes\sigma_y$ (together
with the complex conjugation) in Wootters' 
concurrence formula is that it is a filter constructed
with the comb $\sigma_y C$, rather than the 
time-reversal property of this operator.

{\em Acknowledgments -- }
We would like to thank L.\ Amico, R.\ Fazio, and especially A.\ Uhlmann
for stimulating discussions. This work was supported by the EU 
RTN grant HPRN-CT-2000-00144, the Vigoni Program of the German
Academic Exchange Service and the Sonderforschungsbereich 631 
of the German Research Foundation. J.S. holds a Heisenberg
fellowship from the German Research Foundation.


\begin{thebibliography}{21}
\expandafter\ifx\csname natexlab\endcsname\relax\def\natexlab#1{#1}\fi
\expandafter\ifx\csname bibnamefont\endcsname\relax
  \def\bibnamefont#1{#1}\fi
\expandafter\ifx\csname bibfnamefont\endcsname\relax
  \def\bibfnamefont#1{#1}\fi
\expandafter\ifx\csname citenamefont\endcsname\relax
  \def\citenamefont#1{#1}\fi
\expandafter\ifx\csname url\endcsname\relax
  \def\url#1{\texttt{#1}}\fi
\expandafter\ifx\csname urlprefix\endcsname\relax\def\urlprefix{URL }\fi
\providecommand{\bibinfo}[2]{#2}
\providecommand{\eprint}[2][]{\url{#2}}

\bibitem[{\citenamefont{Bell}(1987)}]{Bell87}
\bibinfo{author}{\bibfnamefont{J.}~\bibnamefont{Bell}},
  \emph{\bibinfo{title}{Speakable and unspeakable in Quantum Mechanics}}
  (\bibinfo{publisher}{Cambridge University Press, Cambridge},
  \bibinfo{year}{1987}).

\bibitem[{\citenamefont{Bennett et~al.}(1996)\citenamefont{Bennett, DiVincenzo,
  Smolin, and Wootters}}]{BennettDiVincenzo96}
\bibinfo{author}{\bibfnamefont{C.~H.} \bibnamefont{Bennett}},
  \bibinfo{author}{\bibfnamefont{D.~P.}~\bibnamefont{DiVincenzo}},
  \bibinfo{author}{\bibfnamefont{J.~A.} \bibnamefont{Smolin}},
  \bibnamefont{and} \bibinfo{author}{\bibfnamefont{W.~K.}
  \bibnamefont{Wootters}}, \bibinfo{journal}{Phys. Rev. A}
  \textbf{\bibinfo{volume}{54}}, \bibinfo{pages}{3824} (\bibinfo{year}{1996}).

\bibitem[{\citenamefont{Hill and Wootters}(1997)}]{Hill97}
\bibinfo{author}{\bibfnamefont{S.}~\bibnamefont{Hill}} \bibnamefont{and}
  \bibinfo{author}{\bibfnamefont{W.~K.} \bibnamefont{Wootters}},
  \bibinfo{journal}{Phys. Rev. Lett.} \textbf{\bibinfo{volume}{78}},
  \bibinfo{pages}{5022} (\bibinfo{year}{1997}).

\bibitem[{\citenamefont{Wootters}(1998)}]{Wootters98}
\bibinfo{author}{\bibfnamefont{W.~K.}~\bibnamefont{Wootters}},
  \bibinfo{journal}{Phys. Rev. Lett.} \textbf{\bibinfo{volume}{80}},
  \bibinfo{pages}{2245} (\bibinfo{year}{1998}).

\bibitem[{\citenamefont{Coffman et~al.}(2000)\citenamefont{Coffman, Kundu, and
  Wootters}}]{Coffman00}
\bibinfo{author}{\bibfnamefont{V.}~\bibnamefont{Coffman}},
  \bibinfo{author}{\bibfnamefont{J.}~\bibnamefont{Kundu}}, \bibnamefont{and}
  \bibinfo{author}{\bibfnamefont{W.~K.} \bibnamefont{Wootters}},
  \bibinfo{journal}{Phys. Rev. A} \textbf{\bibinfo{volume}{61}},
  \bibinfo{pages}{052306} (\bibinfo{year}{2000}).

\bibitem[{\citenamefont{Uhlmann}(2000)}]{Uhlmann}
\bibinfo{author}{\bibfnamefont{A.}~\bibnamefont{Uhlmann}},
  \bibinfo{journal}{Phys. Rev. A} \textbf{\bibinfo{volume}{62}},
  \bibinfo{pages}{032307} (\bibinfo{year}{2000}).

\bibitem[{\citenamefont{D\"ur et~al.}(2000)\citenamefont{D\"ur, Vidal, and
  Cirac}}]{Duer00}
\bibinfo{author}{\bibfnamefont{W.}~\bibnamefont{D\"ur}},
  \bibinfo{author}{\bibfnamefont{G.}~\bibnamefont{Vidal}}, \bibnamefont{and}
  \bibinfo{author}{\bibfnamefont{J.~I.} \bibnamefont{Cirac}},
  \bibinfo{journal}{Phys. Rev. A} \textbf{\bibinfo{volume}{62}},
  \bibinfo{pages}{062314} (\bibinfo{year}{2000}).

\bibitem[{\citenamefont{Vidal}(2000)}]{MONOTONES}
\bibinfo{author}{\bibfnamefont{G.}~\bibnamefont{Vidal}},
  \bibinfo{journal}{J.Mod.Opt.} \textbf{\bibinfo{volume}{47}},
  \bibinfo{pages}{355} (\bibinfo{year}{2000}).

\bibitem[{\citenamefont{Bennett et~al.}(2001)\citenamefont{Bennett, Popescu,
  Rohrlich, Smolin, and Thapliyal}}]{SLOCC}
\bibinfo{author}{\bibfnamefont{C.~H.} \bibnamefont{Bennett}},
  \bibinfo{author}{\bibfnamefont{S.}~\bibnamefont{Popescu}},
  \bibinfo{author}{\bibfnamefont{D.}~\bibnamefont{Rohrlich}},
  \bibinfo{author}{\bibfnamefont{J.~A.} \bibnamefont{Smolin}},
  \bibnamefont{and} \bibinfo{author}{\bibfnamefont{A.~V.}
  \bibnamefont{Thapliyal}}, \bibinfo{journal}{Phys. Rev. A}
  \textbf{\bibinfo{volume}{63}}, \bibinfo{pages}{012307}
  (\bibinfo{year}{2001}).

\bibitem[{\citenamefont{Verstraete et~al.}(2003)\citenamefont{Verstraete,
  Dehaene, and Moor}}]{VerstraeteDM03}
\bibinfo{author}{\bibfnamefont{F.}~\bibnamefont{Verstraete}},
  \bibinfo{author}{\bibfnamefont{J.}~\bibnamefont{Dehaene}}, \bibnamefont{and}
  \bibinfo{author}{\bibfnamefont{B.} \bibnamefont{DeMoor}},
  \bibinfo{journal}{Phys. Rev. A} \textbf{\bibinfo{volume}{68}},
  \bibinfo{pages}{012103} (\bibinfo{year}{2003}).

\bibitem[{\citenamefont{Cereceda}()}]{Cereceda03}
\bibinfo{author}{\bibfnamefont{J.~L.} \bibnamefont{Cereceda}},
  \bibinfo{note}{quant-ph/0305043}.

\bibitem[{\citenamefont{Briand et~al.}(2004)\citenamefont{Briand, Luque,
  Thibon, and Verstraete}}]{Briand03}
\bibinfo{author}{\bibfnamefont{E.}~\bibnamefont{Briand}},
  \bibinfo{author}{\bibfnamefont{J.-G.} \bibnamefont{Luque}},
  \bibinfo{author}{\bibfnamefont{J.-Y.} \bibnamefont{Thibon}},
  \bibnamefont{and}
  \bibinfo{author}{\bibfnamefont{F.}~\bibnamefont{Verstraete}},
  \bibinfo{journal}{J. Math. Phys.} \textbf{\bibinfo{volume}{45}},
  \bibinfo{pages}{4855} (\bibinfo{year}{2004}),
  \bibinfo{note}{quant-ph/0306122}.

\bibitem[{\citenamefont{Barnum et~al.}(2004)\citenamefont{Barnum, Knill, Ortiz,
  Somma, and Viola}}]{Barnum03}
\bibinfo{author}{\bibfnamefont{H.}~\bibnamefont{Barnum}},
  \bibinfo{author}{\bibfnamefont{E.}~\bibnamefont{Knill}},
  \bibinfo{author}{\bibfnamefont{G.}~\bibnamefont{Ortiz}},
  \bibinfo{author}{\bibfnamefont{R.}~\bibnamefont{Somma}}, \bibnamefont{and}
  \bibinfo{author}{\bibfnamefont{L.}~\bibnamefont{Viola}},
  \bibinfo{journal}{Phys. Rev. Lett.} \textbf{\bibinfo{volume}{92}},
  \bibinfo{pages}{107902} (\bibinfo{year}{2004}).

\bibitem[{\citenamefont{Heydari and Bj{\"o}rk}(2004)}]{HEYDARI04}
\bibinfo{author}{\bibfnamefont{H.}~\bibnamefont{Heydari}} \bibnamefont{and}
  \bibinfo{author}{\bibfnamefont{G.}~\bibnamefont{Bj{\"o}rk}},
  \bibinfo{journal}{J. Phys. A} \textbf{\bibinfo{volume}{37}},
  \bibinfo{pages}{9251} (\bibinfo{year}{2004}),
  \bibinfo{note}{quant-ph/0401129}.

\bibitem[{\citenamefont{Meyer and Wallach}()}]{Wallach}
\bibinfo{author}{\bibfnamefont{D.~A.} \bibnamefont{Meyer}} \bibnamefont{and}
  \bibinfo{author}{\bibfnamefont{N.~R.} \bibnamefont{Wallach}},
  \bibinfo{note}{quant-ph/0108104}.

\bibitem[{\citenamefont{Wong and Christensen}(2001)}]{Wong00}
\bibinfo{author}{\bibfnamefont{A.}~\bibnamefont{Wong}} \bibnamefont{and}
  \bibinfo{author}{\bibfnamefont{N.}~\bibnamefont{Christensen}},
  \bibinfo{journal}{Phys. Rev. A} \textbf{\bibinfo{volume}{63}},
  \bibinfo{pages}{044301} (\bibinfo{year}{2001}).

\bibitem[{\citenamefont{Toth and Guehne}(2005)}]{Toth2005}
\bibinfo{author}{\bibfnamefont{G.}~\bibnamefont{Toth}} \bibnamefont{and}
  \bibinfo{author}{\bibfnamefont{O.}~\bibnamefont{Guehne}},
  \bibinfo{journal}{Phys. Rev. Lett.} \textbf{\bibinfo{volume}{94}},
  \bibinfo{pages}{060501} (\bibinfo{year}{2005}),
  \bibinfo{note}{quant-ph/0405165}.

\bibitem[{\citenamefont{Gisin and Bechmann-Pasquinucci}(1998)}]{Gisin98}
\bibinfo{author}{\bibfnamefont{N.}~\bibnamefont{Gisin}} \bibnamefont{and}
  \bibinfo{author}{\bibfnamefont{H.}~\bibnamefont{Bechmann-Pasquinucci}},
  \bibinfo{journal}{Phys. Lett. A} \textbf{\bibinfo{volume}{246}},
  \bibinfo{pages}{1} (\bibinfo{year}{1998}).

\bibitem[{\citenamefont{Verstraete et~al.}(2002)\citenamefont{Verstraete,
  Dehaene, Moor, and Verschelde}}]{VerstraeteDMV02}
\bibinfo{author}{\bibfnamefont{F.}~\bibnamefont{Verstraete}},
  \bibinfo{author}{\bibfnamefont{J.}~\bibnamefont{Dehaene}},
  \bibinfo{author}{\bibfnamefont{B.} \bibnamefont{DeMoor}}, \bibnamefont{and}
  \bibinfo{author}{\bibfnamefont{H.}~\bibnamefont{Verschelde}},
  \bibinfo{journal}{Phys. Rev. A} \textbf{\bibinfo{volume}{65}},
  \bibinfo{pages}{052112} (\bibinfo{year}{2002}).

\bibitem[{\citenamefont{Briand et~al.}(2003)\citenamefont{Briand, Luque, and
  Thibon}}]{BriandLT03}
\bibinfo{author}{\bibfnamefont{E.}~\bibnamefont{Briand}},
  \bibinfo{author}{\bibfnamefont{J.-G.} \bibnamefont{Luque}}, \bibnamefont{and}
  \bibinfo{author}{\bibfnamefont{J.-Y.} \bibnamefont{Thibon}},
  \bibinfo{journal}{J. Phys. A} \textbf{\bibinfo{volume}{36}},
  \bibinfo{pages}{9915} (\bibinfo{year}{2003}),
  \bibinfo{note}{quant-ph/0304026}.

\bibitem[{\citenamefont{Miyake}(2003)}]{Myake03}
\bibinfo{author}{\bibfnamefont{A.}~\bibnamefont{Miyake}},
  \bibinfo{journal}{Phys. Rev. A} \textbf{\bibinfo{volume}{67}},
  \bibinfo{pages}{012108} (\bibinfo{year}{2003}).

\end{thebibliography}

\end{document}